\documentclass[journal]{IEEEtran}

\usepackage{xcolor}
\usepackage{amsmath}
\usepackage{mathrsfs}
\usepackage{amsfonts}
\usepackage{amssymb}
\usepackage{bm}
\usepackage{bbm}
\usepackage{relsize}
\usepackage{mathtools}
\usepackage{algorithm}
\usepackage{algpseudocode}
\usepackage[colorlinks,urlcolor=blue,linkcolor=blue,citecolor=blue,breaklinks=false]{hyperref}
\usepackage{url}
\usepackage{subfigure}

\usepackage[nocompress]{cite}

\ifCLASSINFOpdf
  \usepackage[pdftex]{graphicx}
\else
\fi

\begin{document}
\title{\huge Foldable Antenna Array in Space-Air-Ground Integrated Networks: Architectures and Applications}
%
%
%

\author{Jikang~Deng,~\IEEEmembership{Graduate Student Member,~IEEE,} Ki-Hong~Park,~\IEEEmembership{Senior Member,~IEEE,} and Mohamed-Slim~Alouini,~\IEEEmembership{Fellow,~IEEE}
		
    \thanks{Jikang~Deng, Ki-Hong Park, and Mohamed-Slim~Alouini are with CEMSE Division, King Abdullah University of Science and Technology (KAUST), Thuwal, 23955-6900, Kingdom of Saudi
Arabia (KSA) (email: \href{mailto:jikang.deng@kaust.edu.sa}{jikang.deng@kaust.edu.sa};
\href{mailto:kihong.park@kaust.edu.sa}{kihong.park@kaust.edu.sa}; \href{mailto:slim.alouini@kaust.edu.sa}{slim.alouini@kaust.edu.sa};)}
	}

\maketitle

\begin{abstract}
Space-air-ground integrated networks (SAGINs) integrate heterogeneous platforms with different coverage, mobility, and payload constraints, which creates strong demands for flexible antenna architectures. Foldable antenna arrays (FAAs) provide a promising solution by reconfiguring array geometry, antenna positions, and orientations through controllable joints or hinges. However, existing studies on FAAs have mainly considered simple folding configurations and limited deployment benefits, while FAAs' architectures and broader potential for SAGIN remain largely unexplored. In this article, we present the fundamentals and architectures of FAA, including folding mechanisms, foldability levels, and functional capabilities. We then examine FAA applications for ground base stations, low-altitude platforms, high-altitude platforms, and satellites, together with their potential challenges and solutions. Finally, we highlight open issues and future research directions, and provide a case study to demonstrate the FAA-enabled communication performance improvement in SAGIN.

\end{abstract}

\IEEEpeerreviewmaketitle

\section{Introduction}
\IEEEPARstart{N}{ext-generation} wireless networks are expected to provide high-capacity, reliable, and ubiquitous connectivity. Space-air-ground integrated networks (SAGINs) support this vision by integrating satellites, high-altitude platforms (HAPs), low-altitude platforms (LAPs), and ground base stations (GBSs). They provide complementary coverage but differ considerably in mobility, payload capacity, and operating environment, which requires antenna arrays to support spatially diverse links under platform-specific constraints. Although large-scale multiple-input multiple-output (MIMO) offers substantial array and spatial multiplexing gains, simply increasing the number of antennas cannot fully exploit spatial channel variations with fixed antenna array geometry. This limitation highlights the need for application-aware and environment-aware reconfigurable antenna architectures.

Movable antennas (MAs) and fluid antenna systems reconfigure wireless channels by adjusting antenna positions \cite{zhu2023movable,wong2020fluid}. More recently, six-dimensional movable antennas (6DMAs) have been proposed to jointly control the 3D positions and orientations of antenna surfaces \cite{shao20256dma}, while rotatable antennas (RAs) adjust antenna orientations at fixed locations \cite{zheng2026rotatable}. These architectures provide different spatial degrees of freedom but may require additional movement space or complicated mechanical structures, which should be carefully evaluated when deployed on satellites and other payload-constrained SAGIN platforms.

Foldable antenna array (FAA) provides a complementary approach to spatial channel reconfiguration \cite{wang2026foldable, ning2025movable}. An FAA consists of array segments connected by controllable folding joints or hinges, which allow antenna positions and orientations to be reconfigured through folding. Its feasibility is supported by previous studies on origami-inspired antennas and reflectarrays for CubeSats \cite{park2023shape}, which reduce storage volume during launch and provide a large aperture after deployment. With coordinated folding, the FAA can flexibly reconfigure its array geometry to adapt to different communication conditions and requirements. This flexibility demonstrates its potential for deployment across various SAGIN platforms to improve communication performance.

Motivated by these advantages, this article provides an overview of FAA-enabled SAGIN (FAA-SAGIN). It covers the FAA fundamentals and representative architectures, as well as its applications across different SAGIN platforms. The potential integrations of the FAA system with emerging communication, sensing, and intelligent technologies are also explored in this article.

\section{FAA Fundamentals and Architectures}
\label{section_system_model_FULA}
An FAA consists of multiple rigid subarray segments connected through controllable folding joints or hinges. By adjusting the folding angles of these joints, the FAA can reconfigure the positions and orientations of its antenna elements, thereby reshaping the array geometry. This introduces an additional degree of freedom in the array geometry domain, which enables the FAA to reconfigure wireless channels based on communication requirements and improve communication performance. The representative fundamentals, architectures, and functional capabilities are summarized in Fig.~\ref{Fig_architecture}.

\subsection{Basic Folding Mechanisms}
Based on the dimensionality of the original array, folding mechanisms can be broadly classified into linear and planar folding. 

\begin{figure*}[ht]
  \centering
  \includegraphics[width = 16cm]{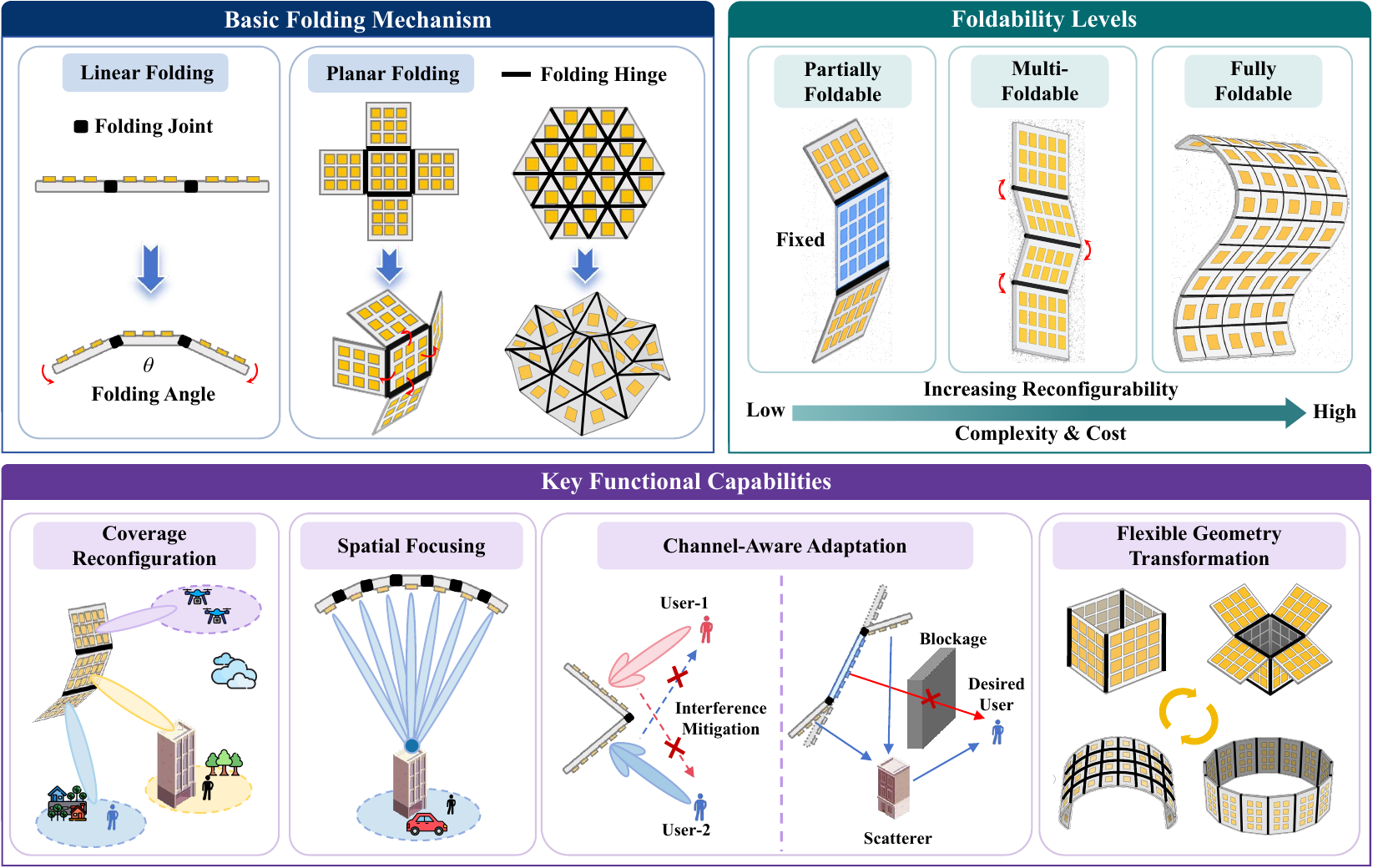}
  \caption{FAA fundamentals, architectures, and key functional capabilities.}
  \label{Fig_architecture}
\end{figure*}

\textbf{Linear Folding:}
Linear folding is mainly applied to linear antenna arrays, where the original array is partitioned into several rigid linear subarray segments connected by folding joints. The folding joint local positions determine the array partitioning and the number of antenna elements in each segment, while the folding angles determine the positions and boresight directions of the folded segments. When multiple joints are deployed on the array, a cascaded folding effect will be observed in all subsequent segments. Consequently, the final position and boresight direction of each antenna element are determined by the cumulative folding angles of the preceding joints.

\textbf{Planar Folding:}
Planar folding extends the same principle to planar antenna arrays, where the original array is partitioned into multiple rigid panels, with rectangular, triangular, or hexagonal shapes. Adjacent panels are connected through hinges deployed in either row or column directions. With folding, the planar array can be transformed into various 3D geometries, as illustrated in Fig.~\ref{Fig_architecture}. 
Planar folding can be implemented using mechanical rotators or deployable origami-based structures. Compared with linear folding, it offers greater geometric flexibility by allowing multiple panels to rotate about different axes. However, the feasible array geometries are usually constrained by the allowable folding angle ranges and mechanical compatibility between adjacent panels. Moreover, since 3D geometries may complicate the RF interconnections across different panels, the greater shape diversity is generally achieved at the cost of increased mechanical and control complexity.

\subsection{Foldability Levels}
According to the density and reconfigurability of the folding joints, FAA architectures can be classified into partially foldable, multi-foldable, and fully foldable arrays, as shown in Fig.~\ref{Fig_architecture}.

\textbf{Partially Foldable Array:}
A partially foldable array consists of fixed and foldable segments. The fixed segments provide stable coverage and can support synchronization and control signaling, while the foldable segments support adaptive coverage enhancement, interference mitigation, and link improvement. This architecture is particularly suitable for scenarios with clustered or relatively stable user distributions, where the foldable segments can be oriented toward high-demand areas to enhance channel capacity.

\textbf{Multi-Foldable Array:}
A multi-foldable array employs multiple controllable joints or hinges without fixed segments, which provide a larger geometric reconfiguration space than a partially foldable array. Through coordinated adjustment of multiple foldable segments, the FAA can form various geometries in response to communication needs. This architecture is therefore well suited to dynamic or spatially dispersed user distributions, where greater geometric flexibility is required to support varying user locations and traffic demands.

\textbf{Fully Foldable Array:}
A fully foldable array deploys a folding joint or hinge between every pair of adjacent antenna elements or smallest rigid panels, thereby providing the highest level of geometric reconfigurability. With densely deployed joints and sufficiently small element spacing, a large-scale linear array can form various curved geometries through coordinated folding. Similarly, a densely hinged planar array can then form cylindrical and other curved or conformal surfaces. Therefore, a fully foldable array can be regarded as a discrete approximation of a continuously bendable antenna array, which makes it mainly suitable for conformal deployment and multifunctional application scenarios.

Among the above architectures, increased reconfigurability generally comes at the cost of higher implementation cost, power consumption, and control complexity. To balance flexibility and complexity, practical FAA designs should therefore pursue the minimum sufficient foldability level, i.e., the smallest number of joints or hinges required to achieve the desired communication performance. Moreover, folding changes the relative positions and orientations of antenna elements. Therefore, the potential mutual coupling, inter-segment signal reflections, and segment collisions in the FAA system should also be carefully modeled and evaluated to ensure reliable FAA operation.

\subsection{Key Functional Capabilities}
As summarized in Fig.~\ref{Fig_architecture}, four representative functional capabilities of FAAs are discussed below.

\textbf{Coverage Reconfiguration:}
For directional antenna elements, the conventional fixed arrays provide fixed coverage and typically require platform reorientation or movement to change the coverage direction. In contrast, an FAA can reconfigure its coverage by adjusting the orientations of individual array segments through folding. Selected segments can redirect their radiation toward users in different directions, thereby improving the sum rate and user fairness.

\textbf{Spatial Focusing:}
By reconfiguring its geometry, an FAA can direct more radiated energy toward the intended receiver or user cluster \cite{zheng2026rotatablesurvey}. Specifically, the array can be folded into a concave arc-shaped geometry so that the boresight directions of multiple directional antenna elements point toward a common spatial region. We refer to this capability as spatial focusing, where the directional antenna gains are jointly enhanced toward the target region to improve the received signal strength.

\textbf{Channel-Aware Adaptation:}
An FAA can adapt its folding configuration to the channel conditions. Specifically, array segments can be oriented to strengthen the link toward the desired user, while the radiation nulls or low-gain directions can be steered toward interfering users. Moreover, with different array segments oriented toward spatially separated user clusters, the FAA can reduce inter-user channel correlation and thus facilitate interference management. In addition, under line-of-sight blockage, selected segments can also be redirected toward dominant scattering paths to establish alternative propagation links and improve communication reliability.

\textbf{Flexible Geometry Transformation:}
A distinctive capability of an FAA is its flexible transformation into different array geometries according to communication requirements. 
It can reproduce conventional array geometries, such as circular and cubic arrays, and then support their corresponding functions, including wide-angle coverage and multi-directional transmission.
Beyond these conventional configurations, the FAA can also form irregular and asymmetric geometries to support complicated propagation environments. By dynamically switching among different geometries, the FAA can effectively adapt to evolving communication demands or environments.

\section{FAA Applications in SAGIN}
\begin{figure*}[ht]
  \centering
  \includegraphics[width = 14cm]{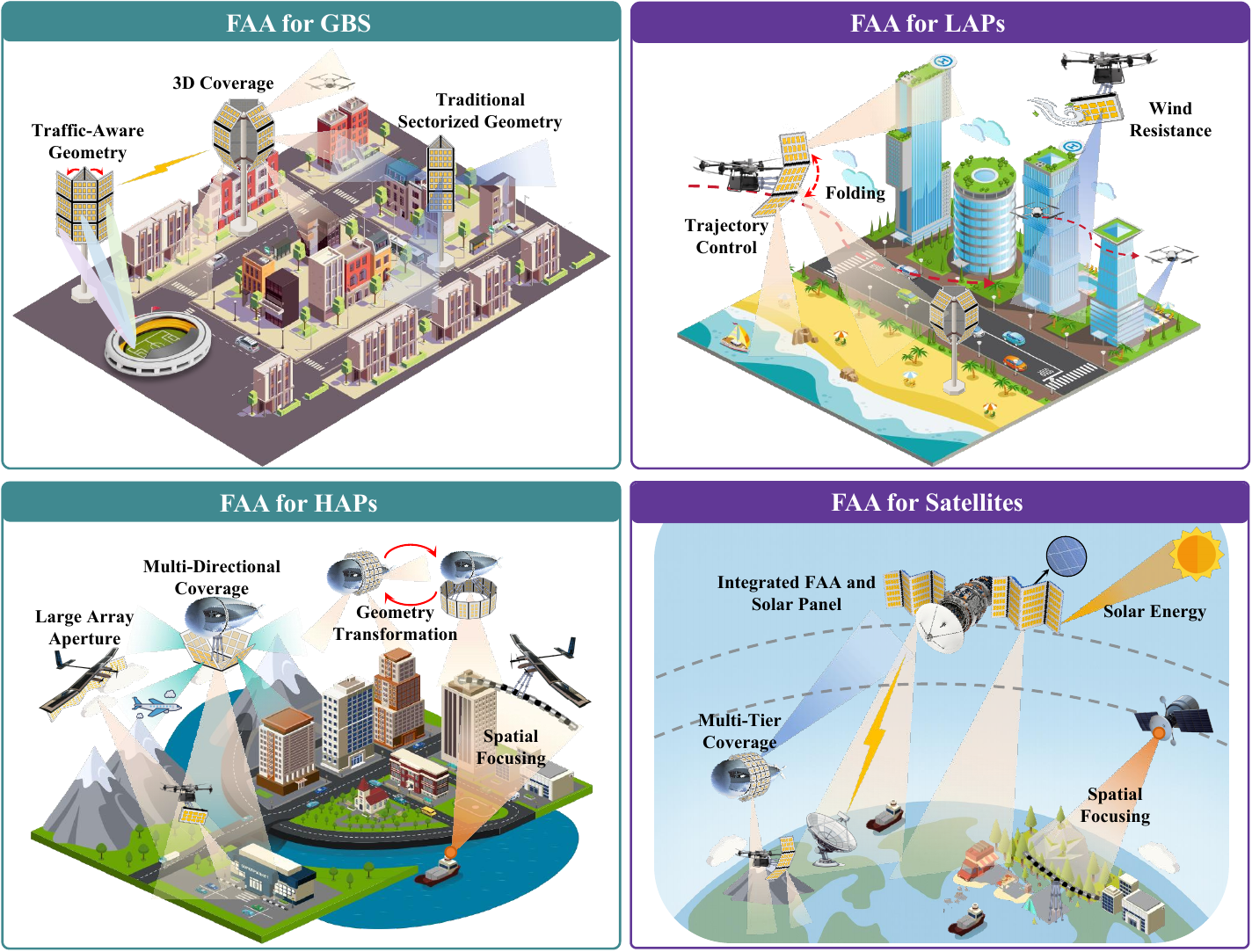}
  \caption{Illustration of FAA-enhanced SAGIN.}
  \vspace{-1em}
  \label{Fig_application_scenario}
\end{figure*}

In this section, we examine the FAA applications across SAGIN platforms and highlight their key benefits and potential challenges, as illustrated in Fig.~\ref{Fig_application_scenario}.

\subsection{Ground Base Stations (GBS)}
Conventional GBSs are typically deployed on elevated sites, such as buildings or hills, and benefit from reliable power supply, stable backhaul connectivity, and sufficient payload capacity. They commonly employ fixed three-sector antenna arrays with downtilt configurations optimized primarily for 2D terrestrial coverage \cite{yang2025flexible}. Consequently, aerial users are often served through the upper array sidelobes, resulting in fragmented coverage and strong inter-cell interference. To address this limitation, an FAA can be deployed in each sector to enhance the 3D coverage capability of GBSs. Specifically, different array segments can be oriented toward uptilt, vertical, and downtilt directions to simultaneously serve aerial users, neighboring GBSs, and ground users, respectively. This allows targets at different altitudes to be covered by the main lobe of dedicated array segments. Moreover, such segment-wise orientation can also improve the spatial separation among transmission links and mitigate interference.

Furthermore, an FAA can also adapt its geometry to maintain reliable communication under uneven spatial communication traffic conditions. In conventional three-sector arrays, traffic demands may vary significantly across sectors, leading to inefficient utilization of antenna resources.
In contrast, an FAA-based GBS can redirect the underutilized arrays of lightly loaded sectors to serve regions with high traffic demand. When traffic becomes balanced, the FAA can return to the standard three-sector configuration. This traffic-aware coverage adaptation helps improve the antenna resource utilization efficiency.

\subsection{Low-Altitude Platforms (LAPs)}
LAPs mainly include uncrewed aerial vehicles (UAVs) and electric vertical take-off and landing (eVTOL) aircraft. They are expected to play an important role in 6G networks, particularly for emergency response and temporary service enhancement. They can also support the emerging low-altitude economy (LAE) through flexible aerial connectivity for transportation and logistics. In the following discussion, we focus on UAVs as representative LAPs.
Mounting an FAA on a UAV enhances the spatial flexibility by jointly exploiting UAV 3D mobility and array geometry reconfiguration. Since the UAV trajectory and FAA folding configuration typically affect the channel conditions over different timescales, their joint design naturally leads to two-timescale optimization problems. Moreover, the FAA-mounted UAV can also adapt its spatial coverage to terrestrial and aerial users at different altitudes. In addition, for UAV swarms, different array segments can be oriented to facilitate inter-UAV cooperation and mitigate interference.

UAVs usually have stringent payload capacity and energy supply constraints, which limit the aperture size and power consumption of the mounted FAA. For example, the DJI Matrice 350 RTK UAV platform has a payload capacity of 2.7 kg and a flight time of approximately 31 minutes. 
Nevertheless, it is worth noting that a lightweight FAA may also help reduce UAV energy consumption and extend service endurance. 
Specifically, FAA folding can provide part of the channel improvement previously brought by UAV repositioning. Since folding a lightweight FAA generally consumes substantially less energy than UAV propulsion, array folding can substitute for part of the energy-intensive UAV movement, thereby reducing the overall energy consumption.
Additionally, wind-induced aerodynamic loads may affect flight stability \cite{ning2025movable}. Under strong wind conditions, the FAA can be folded into a more compact geometry to reduce wind loads and improve flight stability.

\vspace{-0.5em}
\subsection{High-Altitude Platforms (HAPs)}
HAPs typically operate in the stratosphere at altitudes of around 20 km and can remain airborne for several months. Compared with LAPs, HAPs offer larger payload capacity, sufficient power supply, and wider coverage \cite{deng2026distributed}. For example, the HAP of Stratospheric Platforms Ltd supports a 140 kg maximum load and provides a 20 kW power supply. Although HAPs exhibit less flexibility than UAVs, they are well suited for large-scale FAA deployment.

A large-scale FAA can be reconfigured into various 3D geometries. For instance, a cubic array can provide near-full-space coverage while retaining the simple structure of planar arrays. Moreover, its side panels can be independently folded to adapt the coverage toward multiple spatial directions.
In contrast, a large-scale arc-shaped array can achieve spatial focusing with higher antenna gain and narrower beamwidth, which helps compensate for the severe path loss of HAP links. 
These geometries provide complementary capabilities in coverage and directivity and can be selected or combined according to communication requirements. 
For example, when serving both airliners and ground users, some array segments can form arc-shaped subarrays to track airliners, while the remaining segments maintain coverage toward ground users. 
In addition, FAAs are also promising for HAP-based radar imaging. By orienting different array segments toward the target region, the effective sensing aperture and angular diversity can be increased, which helps improve spatial resolution and imaging quality.

\vspace{-0.4em}
\subsection{Satellites}
Satellites are commonly classified as low Earth orbit (LEO), medium Earth orbit (MEO), and geostationary Earth orbit (GEO) satellites. Among them, LEO satellites are particularly attractive for 6G communications due to their relatively low propagation delay and path loss. However, their limited payload capacity and installation area constrain the size of onboard FAAs.
One promising solution is to integrate FAAs with deployable solar panels. The feasibility of such integration has been demonstrated by NASA's Integrated Solar Array and Reflectarray Antenna (ISARA) mission. In this mission, a high-gain Ka-band reflectarray integrated with solar panels enabled a CubeSat downlink rate of over 100 Mbps \cite{nasa2026communications}. 

Inspired by ISARA, FAA elements can be deployed on the reverse side of solar panels to enlarge the effective aperture without additional deployment structures. To make this integrated FAA–solar-panel structure retain the functional capabilities mentioned earlier, the foldability level of the traditional solar panels should be designed accordingly. For example, with sufficient foldability, the integrated structure can achieve spatial focusing and then improve the link budget for satellite-to-ground or inter-satellite communications. A similar design can also be applied to solar-powered HAPs. 
However, high launch costs and limited in-orbit maintenance require folding joints and hinges to be highly reliable and energy-efficient under harsh space conditions.
Additionally, the FAA configuration and solar panel orientation should be jointly optimized to balance communication and energy harvesting along the satellite orbit. Solar energy harvesting can be prioritized during sunlit periods, while communication is prioritized during eclipse periods.

\section{Open Issues and Future Directions}
This section discusses key open issues and promising research directions for FAA-enabled SAGIN, with the goal of further improving its performance and practical feasibility.

\vspace{-0.2em}
\subsection{Integrated Sensing and Communication (ISAC) in FAA-SAGIN}
ISAC enables SAGIN platforms to jointly support communication and sensing using shared waveform or hardware resources. The FAA's segmented structure creates new opportunities for ISAC in SAGIN. The FAA with folding can create diverse observation angles and propagation paths, and then enhance the target detection accuracy and resolution. In multi-layer SAGIN, different segments may further observe different spatial regions or jointly sense the same target from multiple directions. Meanwhile, the user distributions and traffic demands can be estimated based on sensing results and then used to guide FAA geometry reconfiguration design. This creates a closed-loop and coupled mechanism, where sensing provides environmental information, and the FAA accordingly reshapes its geometry to enhance both communication and sensing performance.
In addition, the segmented FAA provides additional flexibility in allocating and orienting array segments for different functions. Specifically, different FAA segments can be assigned to communication functions, such as access and backhaul, or sensing functions, such as imaging and localization. However, assigning segments to different tasks reduces the effective aperture available to each function. Therefore, efficient segment allocation and geometry optimization require further investigation.

\vspace{-0.3em}
\subsection{Hybrid Reconfigurable Antenna in FAA-SAGIN}
\begin{figure}[ht]
  \centering
  \includegraphics[width = 8.7cm]{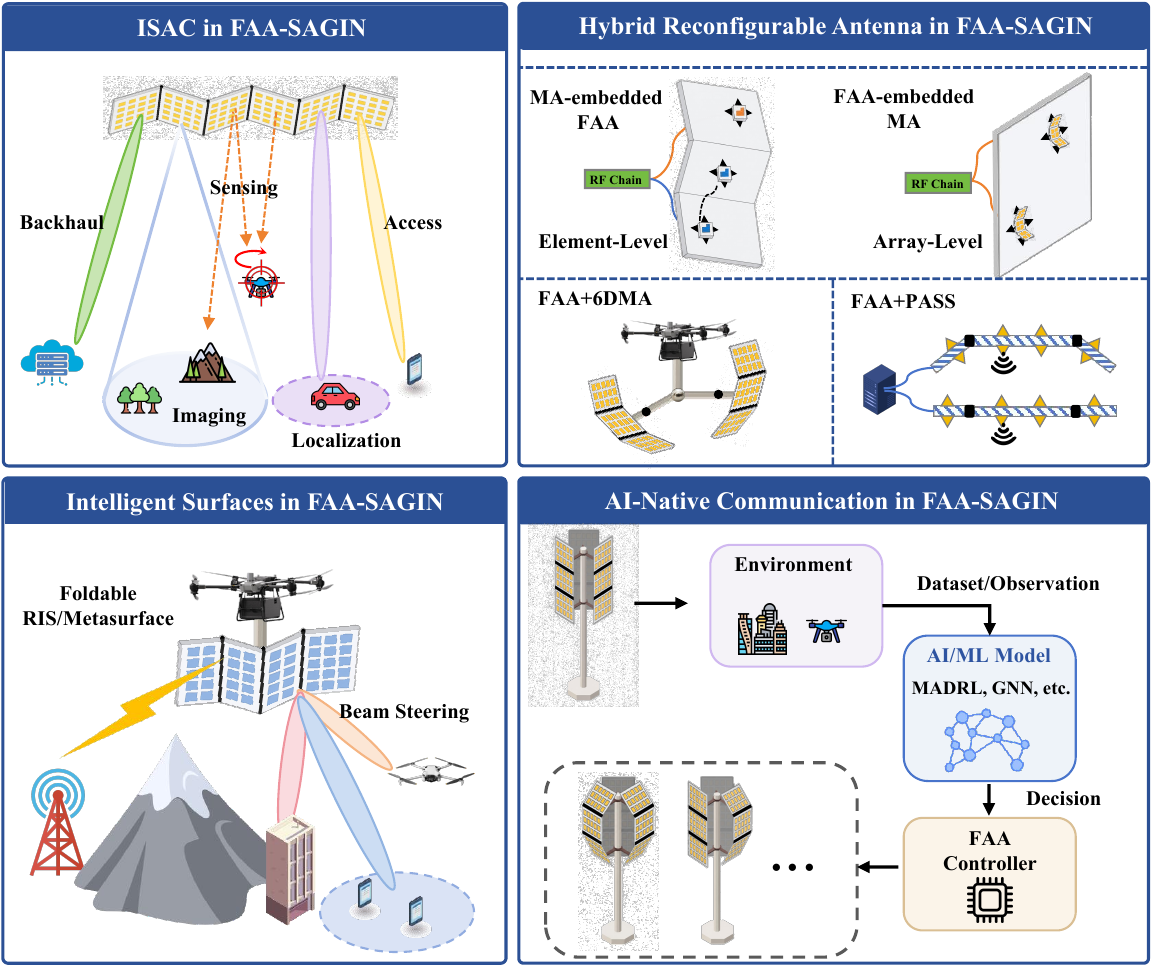}
  \caption{Open issues and future directions for FAA-SAGIN.}
  \vspace{-0.8em}
  \label{Fig_open_issue}
\end{figure}

Movable and reconfigurable antennas offer spatial degrees of freedom that are complementary to those of FAAs, making their integration promising for SAGIN. As illustrated in Fig.~\ref{Fig_open_issue}, such hybrid designs can be implemented at different levels.
At the element level, the MA elements can be deployed on a foldable panel, which provides a larger and more flexible moving region. 
At the array level, an entire FAA can be integrated with MA, 6DMA, or RA mechanisms. This enables both global position/orientation adjustment and internal array geometry reconfiguration through folding. Such a hybrid design is particularly suitable for GBSs and HAPs due to their sufficient installation space and payload capacity. 
It can also support UAV communications by combining large-scale platform movement with fine-grained array position and geometry reconfiguration. However, the resulting system weight, power consumption, and mechanical complexity should be carefully controlled.
Additionally, the folding concept may also be extended to pinching antenna systems (PASS), where foldable waveguides could provide a more flexible deployment region for pinching antennas \cite{liu2025pinching}. However, the feasibility of waveguide folding, together with reliable mechanical and electromagnetic connections between waveguide segments, still requires further investigation.

\subsection{Intelligent Surfaces in FAA-SAGIN}
The folding concept can be extended from active antenna arrays to reconfigurable intelligent surfaces (RISs) and metasurfaces. Metasurface tiles can be mounted on foldable panels or integrated into part of the antenna array aperture. Folding adjusts the positions and orientations of surface segments, while programmable elements control their electromagnetic responses \cite{song2024origami}. 
By adjusting the panel geometry, different surface segments can be better aligned with dominant incident and reflected directions. Moreover, folding and electromagnetic reconfiguration operate at different timescales: the folding adapts to slowly varying large-scale conditions, while the electromagnetic reconfiguration tracks faster channel variations. This motivates joint optimization of the panel geometry and RIS responses rather than optimizing RIS coefficients under a fixed geometry. However, since folding may affect the mutual coupling, polarization, and phase responses of metasurface elements, it is essential to provide accurate electromagnetic modeling and reliable control circuits in this integrated system.

\subsection{AI-Native Communication in FAA-SAGIN}
Artificial intelligence-native (AI-Native) communication is attracting growing interest for intelligent network control. In FAA-SAGIN, AI can support dynamic array geometry optimization and distributed coordination across various platforms \cite{deng2026distributed}. For example, multi-agent deep reinforcement learning (MADRL) can model each folding joint as an agent, which adjusts its position and angle according to channel conditions while avoiding collisions and mitigating interference. For links with FAAs at both ends, coordinated control among heterogeneous SAGIN platforms may require further investigation. Graph neural networks (GNNs) are also well suited to the structural characteristics of FAAs, with joints and connected segments modeled as graph nodes and edges. Through message passing, they can capture geometric coupling among adjacent joints and then generate suitable FAA configurations. 
In addition, for platforms with limited onboard computing resources, TinyML may be employed to enable low-complexity local inference. When computing resources are located remotely from the FAA-mounted platform, the resulting propagation delays will lead to outdated FAA configurations. Their impact and corresponding mitigation strategies should therefore be carefully investigated. 

\section{Case study}
In this section, we evaluate the communication performance of the FAA-enabled SAGIN. 
We consider a downlink system where a HAP at an altitude of 20 km is equipped with a 33-element linear double-joint FAA with directional antenna elements. We assume one joint ($J_1$ or $J_2$) is deployed on each half of the FAA and can only be placed between two adjacent elements. The HAP serves a total of 11 single-antenna multi-tier users, including 3 airplane users, 3 UAV users, and 5 ground users. Several scatterers are included, resulting in Rician channels. The carrier frequency is 30 GHz, the default maximum transmit power is 30 dBm, and the noise power is -120 dBm. We aim to maximize the minimum signal-to-interference-plus-noise ratio (SINR) by jointly optimizing the folding configuration, beamforming, and power allocation.

To solve this problem, we propose a two-layer optimization framework. In the outer layer, a mixed-variable particle swarm optimization (MVPSO) algorithm searches for the FAA folding configuration, which consists of continuous folding angles and discrete folding joint local positions. For each candidate folding configuration, the inner layer jointly optimizes the beamforming and power allocation, and the maximized min-SINR (max--min SINR) is used as the fitness metric for the outer-layer search. Based on this framework, we compare FAA-Optimal, which jointly optimizes the folding configuration, beamforming, and power allocation, with three traditional fixed array (TFA) benchmarks: TFA-ZF and TFA-MRT, which employ zero-forcing (ZF) and maximum-ratio transmission (MRT) beamforming with optimized power allocation, respectively, and TFA-Optimal, which jointly optimizes the beamforming and power allocation.

\begin{figure}[ht]
  \centering
  \includegraphics[width = 8.9cm]{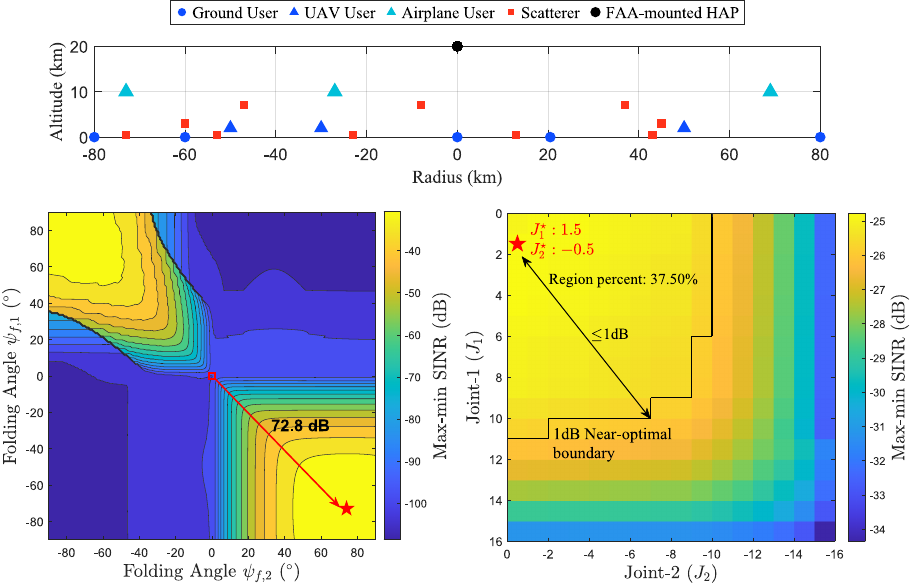}
  \caption{HAP-based communication scenario and max–min SINR under different FAA folding configurations. (Top: user distribution; bottom-left: folding angle heatmap; bottom-right: folding joint local position heatmap)}
  \label{Fig_heatmap}
\end{figure}
Before presenting the performance of the proposed solution, we illustrate the considered HAP-based communication scenario with multi-tier users in Fig.~\ref{Fig_heatmap}. The two heatmaps are obtained through exhaustive search over the folding angles and joint local positions, respectively.
The bottom-left heatmap highlights the benefit of optimizing the folding angles. Compared with the unfolded configuration, $[0^\circ,0^\circ]$, the optimized angles of $[-73^\circ,74^\circ]$ improve the max--min SINR by approximately $72.8$ dB.
The bottom-right heatmap identifies the optimal joint local positions as $J_1^\star=1.5$ and $J_2^\star=-0.5$. Moreover, $37.5\%$ of the evaluated joint local position pairs achieve performance within $1$ dB of the optimum, which indicates the flexibility in selecting joint local positions while maintaining near-optimal performance. These results collectively demonstrate the importance of properly configuring both the folding angles and joint local positions.

\begin{figure}[ht]
  \centering
  \includegraphics[width = 6.7cm]{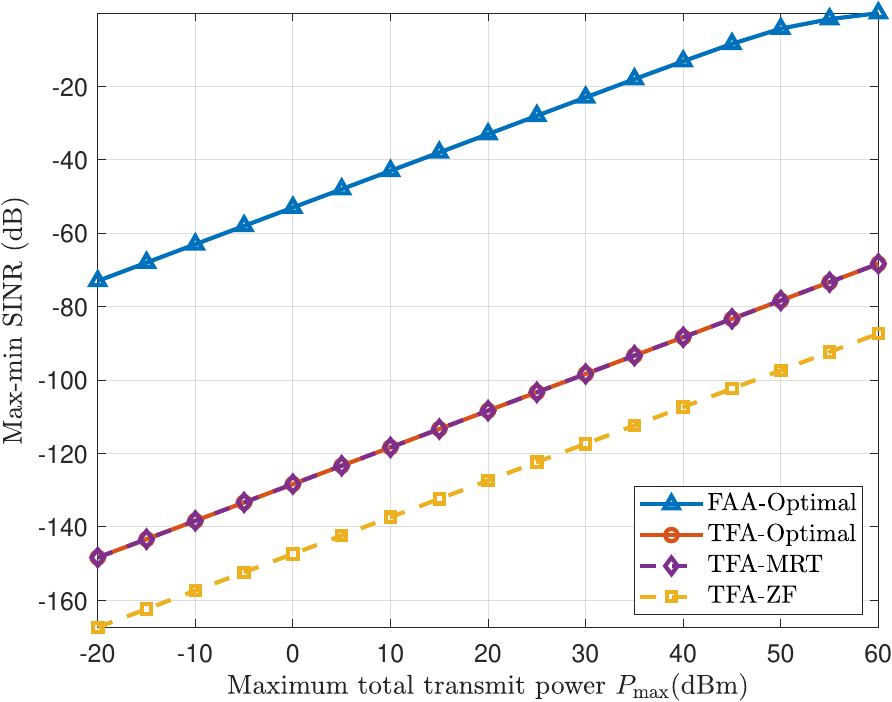}
  \caption{Max–min SINR versus maximum transmit power for FAA and TFA systems}
  \label{Fig_power_comparison}
\end{figure}
Based on the proposed two-layer optimization framework, Fig.~\ref{Fig_power_comparison} shows the max--min SINR performance of FAA and TFA systems versus the maximum transmit power. With the optimized folding configuration, the FAA consistently outperforms the TFA benchmarks across the entire transmit power range. These results demonstrate the effectiveness of the proposed optimization framework and highlight the potential of FAA systems for enhancing communication performance in SAGIN.

\section{Conclusion}\label{section_conclusion}
In this article, we investigated FAA architectures and their applications in SAGIN through adaptive coverage and flexible geometry reconfiguration. Compared with conventional fixed antenna arrays, FAAs can adjust antenna positions and boresight directions to better adapt to channel conditions and improve communication performance. However, practical implementation, deployment, and efficient folding optimization of FAAs in SAGIN remain important challenges. Further research is thus needed to address these issues and facilitate the practical development of FAA-enabled SAGIN.

%





\ifCLASSOPTIONcaptionsoff
  \newpage
\fi



\bibliographystyle{IEEEtran}
\bibliography{IEEEabrv,ref}
\end{document}